**Commentary on: "Unconventional elasticity in smectic-*A* elastomers" by O. Stenull and T. C. Lubensky (0706.3020v1)**

Dominic Kramer, Heino Finkelmann

Institut für Makromolekulare Chemie

Albert Ludwigs-Universität Freiburg, 79104 Freiburg, Germany

Abstract

The reorientation behaviour of a smectic-*A* (*SmA*) elastomer deformed parallel to the smectic layer normal has been interpreted as a *Sm-C* like transition by Stenull and Lubensky. Experiments, however, prove that such a transition does not occur.

Stenull and Lubensky present a description of the elasticity of smectic-*A* (*SmA*) elastomers upon mechanical deformation along the layer normal. They compare their interesting theoretical findings with the experimental data given by Nishikawa some years ago.[1] Upon deformation parallel to the layer normal, the macroscopically ordered network initially exhibits a large modulus which decreases significantly after a threshold strain of about 3%. X-ray measurements show that the smectic layers reorient, indicated by a splitting of the small angle reflections and a loss of intensity. Stenull and Lubensky interpret these observations as a transition to a *SmC*-like state in which the director is oriented parallel to the stress axis, which differs from recent theoretical models [2,3]. However, they do not take into consideration some important experimental observations that cannot be explained by a *SmC*-ordering. On deformation, the sample turns instantly opaque, which is not expected for the proposed *SmC* chevron structure. The order parameter decreases from $S = 0.73$ at $\lambda = 1.00$ to $S = 0.43$ at $\lambda = 1.50$ and the layer distance remains constant throughout this process. At the threshold strain, the modulus decreases from $10^7$ Pa, which reflects the layer compression modulus of the smectic phase, to $10^5$ Pa, a value typical for the corresponding isotropic rubber. For a *SmC*-like network the positional ordering should contribute to the elastic moduli, which is expected to be > $10^5$ Pa. Instead of a *SmC*-ordering, these arguments suggest a breakdown of smectic layering, by a layer rotation and/or a transition to a distorted nematic-like state.

*SmA* elastomers favour a constant layer spacing, so that a deformation along the layer normal will produce shear strains above a crictical stress. Stenull and Lubensky assume that this shear strain causes the *SmC*-ordering. We present shear experiments (Fig.1), which clearly show that a shear deformation has no influence on the *SmA* phase structure. The angle between small and wide angle reflexions stays at a constant value of 90°.

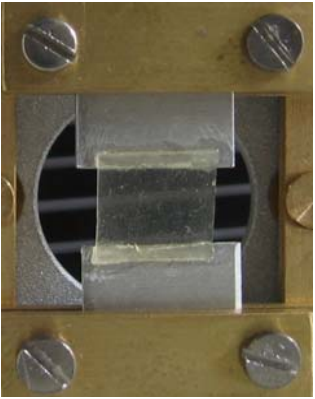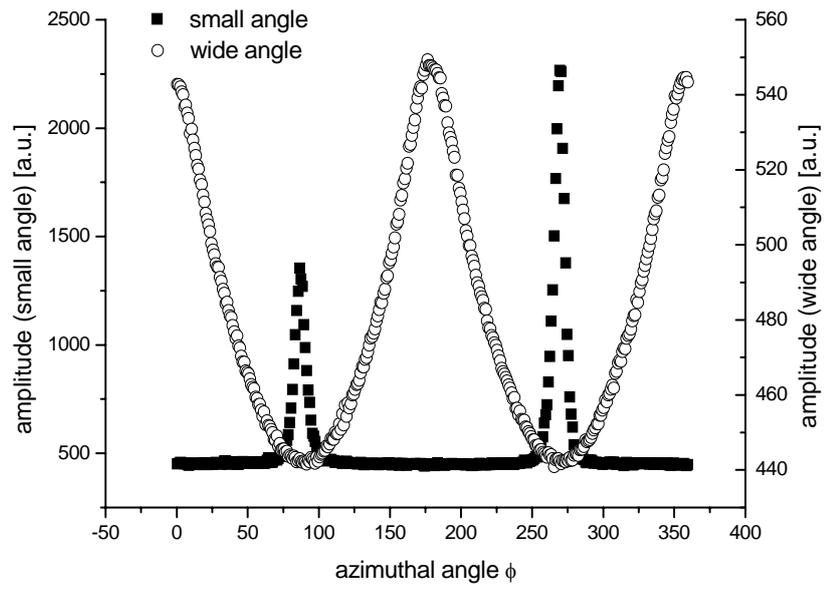

**Figure 1a. Experimental setup and azimuthal intensity distribution.**

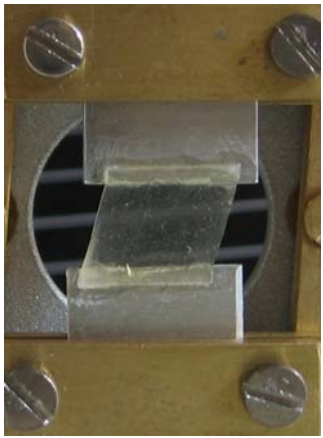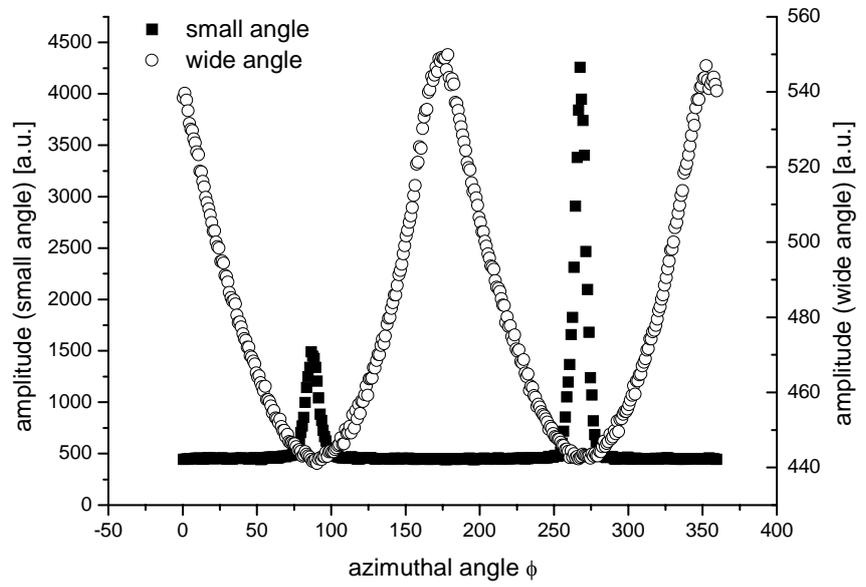

**Figure 1b. Experimental setup and azimuthal intensity after a shear angle of ≈ 14°.**